\documentclass[aps,twoside,twocolumn,floatfix,showpacs,superscriptaddress]{revtex4}
\usepackage{graphics,amssymb,amsmath}

\begin{document}
\title{$1/f$ Spectrum in the Information Transfer Model for Mass Extinction}
\author{B.-G. \surname{YOON}}
\affiliation{Department of Physics, University of Ulsan, Ulsan
680-749, Korea}
\author{M.S. \surname{CHUNG}}
\affiliation{Department of Physics, University of Ulsan, Ulsan
680-749, Korea}
\author{M.Y. \surname{CHOI}}
\affiliation{Department of Physics, Seoul National University,
Seoul 151-747, Korea}
\affiliation{Korea Institute for Advanced Study,
Seoul 130-722, Korea}

\begin{abstract}
We study the information transfer model for biological evolution with several kinds of fitness 
function.  The system is stimulated to evolve into a stationary state, depending on the fitness 
function and on the dimension of the lattice formed by the species.  
In particular the system yields time series of the mutation rate which exhibits the $1/f$ spectrum,
thus explains the power-law behavior in fossil data. 
Effects of shortcuts introduced on the lattice are also examined and the evolution activity is usually found 
to increase in a well-growing system although the reduction of the overall activity may also be observed
in the presence of shortcuts, depending on the initial configuration. 
\end{abstract}


\maketitle

\section{Introduction}
Although it is usually believed that some extinction events of largest scale
were caused by exogenous stresses such as the impact of asteroid,
also observed are background extinctions of all scales~\cite{futuyma,NewPal}. 
Recently, there have been attempts to explain the extinction data in terms of scale invariance~\cite{gisiger}. 
It is argued that the ecosystem is more or less critical, namely, the constituents are correlated in all scales.
As a result of this, the distribution $D(s)$ of extinction events of size $s$ fits to a power law, 
$D(s) \sim s^{-\tau}$ with $1\lesssim \tau \lesssim 2$~\cite{bakpac}. 
In addition, the lifetime distributions of families and genera also display similar power-law 
behavior~\cite{burlando}.  The origin of this apparent criticality is not yet clearly resolved, for
we do not know the dynamics of evolution sufficiently. 
Among many efforts to explain the extinctions of species as natural phenomena without resorting to external
causes, there are the fitness landscape models~\cite{nk}, self-organized critical models~\cite{bak96},
inter-species connection models~\cite{inter}, and environmental stress models~\cite{environ}. 
All of these can explain certain features of the fossil data, although some models may give more 
satisfactory exponents of the power laws than others.  The successes and shortcomings of these models 
are reviewed in Refs.~\onlinecite{NewPal} and \onlinecite{drossel}. 

Several years ago it was suggested that the extinction data can be explained with 
the ``information transfer  model'', in which entropy plays a decisive role for acceptance of 
a mutation event~\cite{choi}.  It takes into consideration the fact that mutations are random 
on the molecular level whereas natural selections are performed via information transfer. 
Interestingly, the resulting dynamics has turned out to be the entropic sampling algorithm, 
which has been successfully applied to the traveling salesman problem~\cite{lee}. 
Accordingly, the information transfer model does not employ the extremal dynamics, 
which evolves the system by sequentially updating or mutating the species with the globally minimum 
value of fitness \cite{BS}.  Nevertheless, if we allow the system to evolve in 
such a direction as to decrease the entropy of the system, power-law behavior emerges on the phenotypic 
level.  This suggests that entropy indeed plays a key role in biological evolution and the 
information transfer may be the origin of self-organized criticality. 
However, there still lacks direct evidence for self-organized criticality;
in particular it has not been addressed whether the mutation rate displays a power-law spectrum. 
Further, it has not been examined how the behavior of the model depends on such ingredients as
the fitness function, the dimension, and the presence of shortcuts. 
This work studies the detailed behavior of the information transfer model: 
the stationary patterns of the mutation rate $R(t)$ and the total fitness $F(t)$, 
as well as the extinction event distribution $D(s)$, depending on the fitness function of species 
and the dimension of the lattice.  It is revealed that the mutation rate displays the $1/f$ 
spectrum, confirming the criticality of the model.  In consideration of the small-world character, 
we also add shortcuts to the lattice and examine their effects on the evolution activity. 

\section{Information Transfer Model}
For completeness, we briefly describe the information transfer model for biological evolution, 
which is essentially the Monte Carlo simulation process adopting the entropic sampling algorithm. 
Consider an ecosystem of $N$ species, in interaction with the environment. 
The configuration of the ecosystem is described by $x \equiv \{x_i\}$, where 
$x_i$ represents the configuration of the $i$-th species and is taken to be a number between zero and unity 
($0\leq x_i < 1$). 
We assume that the fitness of the $i$th species is given by a real function $f_i (x)$, 
which may depend upon the configuration of its neighbors as well as its own configuration $x_i$. 
The total fitness $F(x)$ is then defined to be the sum of the fitness values
of all species in the system:
\begin{equation}
  F(x)=\sum_{i=1}^N f_i(x).
\end{equation}

The information-theoretical entropy $S$ can be obtained as a function of $F$. 
We initialize the ecosystem and make it evolve according to the following rule:
(i) Choose randomly a single species in the system and mutate the chosen species only. 
Denote $x'$ to be the new configuration.
(ii) We compute the entropy change $\Delta S \equiv S(F(x'))-S(F(x))$ 
after the trial mutation $x \rightarrow x'$, and accept the mutation when $\Delta S < 0$; for $\Delta S >0$,
the mutation is accepted with probability $\exp (-\Delta S)$.

In this algorithm, the entropy of the ecosystem is estimated as follows:
Initially the entropy $S(F)$ is set equal to zero for all values of the total fitness $F$. 
We then obtain the histogram $H(F)$ of the total fitness for a short run,
which gives new estimation of $S(F)$:
\begin{equation}
 S(F)= 
 \left\{
  \begin{array}{lr}
  S(F)& \mbox{for $H(F)=0$} \\
  S(F)+\ln H(F)& \mbox{otherwise}.
  \end{array}
 \right.
\end{equation}
In this model, the entropy of the system increases if the system stays long in configurations 
with the same value of $F$. Accordingly, the probability for the mutation to make the system escape 
from a region of the same $F$ is higher than that for the mutation to have the system stay in that region. 
Accordingly, the total fitness $F$ oscillates in time, unless the system falls in a region of the
lowest entropy at the extrema of $F(t)$ and has difficulty in escaping from it. 

For numerical work, $N= L^2 $ species are arranged to form either a circular chain or a square array 
with linear size $L$.  We mainly use $N=16^2$ in the numerical work here. 
Three kinds of fitness function $f_i (x)$ are used: One is the Bak-Sneppen (BS) type fitness function 
$f_i = x_i$~\cite{BS}, which was also employed in Ref.~\onlinecite{choi}. 
The other two are given by 
\begin{equation}\label{xx}
f_i= {\sum_j}'  x_i x_j /2 
\end{equation}
and
\begin{equation}\label{x-x}
f_i= {\sum_j}' {(x_i - x_j )}^2 /2,
\end{equation}
respectively, where the prime restricts the summation to nearest neighbors of the $i$-th species. 
Without knowledge of the relevant coupling form, we just consider these fitness functions 
representing simplest nontrivial couplings between species: 
While the fitness given Eq.~(\ref{xx}) gets large for $x_i$'s of neighboring species both 
large, that in Eq.~(\ref{x-x}) grows with the difference in $x_i$'s between neighboring species.

We initialize the system by assigning a random number in the interval $[1,0)$ to each $x_i$. 
As the system evolves according to the entropic sampling algorithm, 
we obtain the time series of the total fitness $F(t)$ and the mutation rate $R(t)$, 
where the unit of time $t$ is taken to be one Monte Carlo step per species (MCS). 
The mutation rate $R(t)$ is defined to be the number of mutations taking place between 
time $t{-}10$ and $t$, where the time interval of ten (instead of unity) is chosen just
for convenience, in view of the large time scale in fossil data. 
It is indeed confirmed that the overall behavior of the time series does not depend on the length of 
the time interval in which $R(t)$ is sampled. 
Usually we obtain results for an ensemble of $20$ to $30$ sets of species 
with the same lattice type and fitness, and check any dependence of the stationary patterns 
on the initial configuration.

\section{$1/f$ Spectrum and Criticality}

It was observed that the avalanche size distribution $D(s)$ and the waiting time distribution 
$\tilde {D}(t_w )$ satisfy power laws, with the exponents close to two~\cite{choi}.  
An avalanche of size $s$ consists of $s$ mutations, each of which is accepted 
within $\Delta$ trials after the previous mutation. 
In this work we take $\Delta=10$, again considering the large time scale in fossil data, 
and confirm that the resulting power-law behavior is persistent with the exponent not 
much changing as $\Delta$ is varied. 
The waiting time $t_w$ is defined to be 
the time interval for a certain species or for a species of maximum fitness 
to wait until the next mutation occurs. 
It was then claimed that the obtained exponents were compatible with the fossil data. 
If one of the two distributions satisfies a power law, the other also satisfies a power law 
with a similar exponent.  Hence we present here only the results of the avalanche size distribution $D(s)$.

\begin{figure}
\centering{\resizebox*{!}{10cm}{\includegraphics{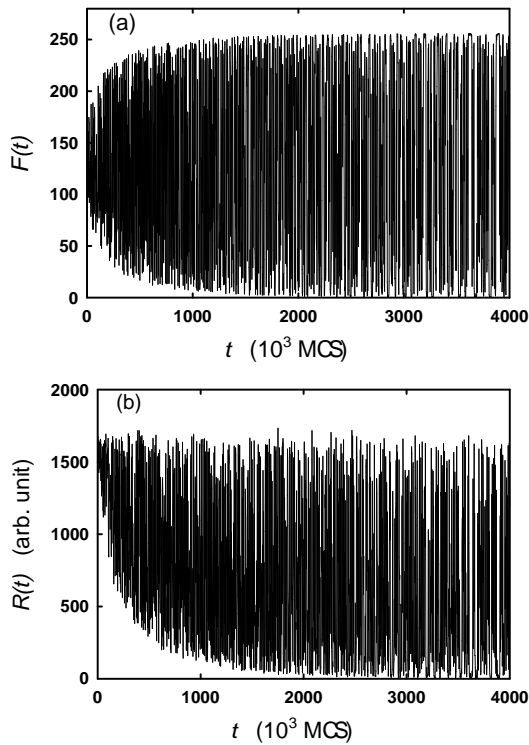}}}
\caption[0]{Coarse-grained time series in a system with the fitness function $f_i = x_i$: 
(a) total fitness $F(t)$ and (b) mutation rate $R(t)$.}
\label{fig:frbs}
\end{figure}
\begin{figure}[b!]
\centering{\resizebox*{!}{10cm}{\includegraphics{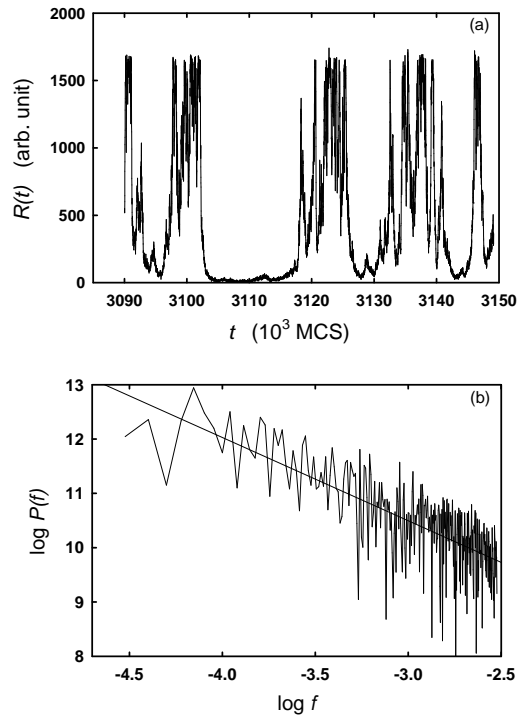}}}
\caption[0]{(a) Part of the time series $R(t)$ in the fine time scale and (b) power
spectrum $P(f)$ of the time series data in Fig.~\ref{fig:frbs}, exhibiting the power-law
behavior $f^{-\alpha}$. The straight line corresponds to the exponent $\alpha =1.5$. 
}
\label{fig:rft}
\end{figure}

We begin with the BS type fitness function: $f_i =x_i$ and show in Fig.~\ref{fig:frbs} 
long-term behaviors of the (coarse-grained) time series $F(t)$ and $R(t)$. 
At earlier time, both $F(t)$ and $R(t)$ oscillate irregularly while the average amplitudes 
keep increasing.  Eventually, the system grows into a stationary state, irrespective of 
the initial configuration.  In this regime, the time series $R(t)$ is shown in detail in 
Fig.~\ref{fig:rft}(a), characterized by aperiodic repetitions of relatively dormant periods. 
Remarkably, the power spectrum $P(f)$, computed directly from the time series data $R(t)$,
exhibits the power-law behavior $f^{-\alpha}$ in frequency $f$, 
with the exponent $\alpha \approx 1.5$, as shown in Fig.~\ref{fig:rft}(b). 
It is pleasing that this $1/f$ behavior of the power spectrum is indeed consistent with that 
observed in the extinction data~\cite{lovf,PS}. 

\begin{figure}
\centering{\resizebox*{!}{6cm}{\includegraphics{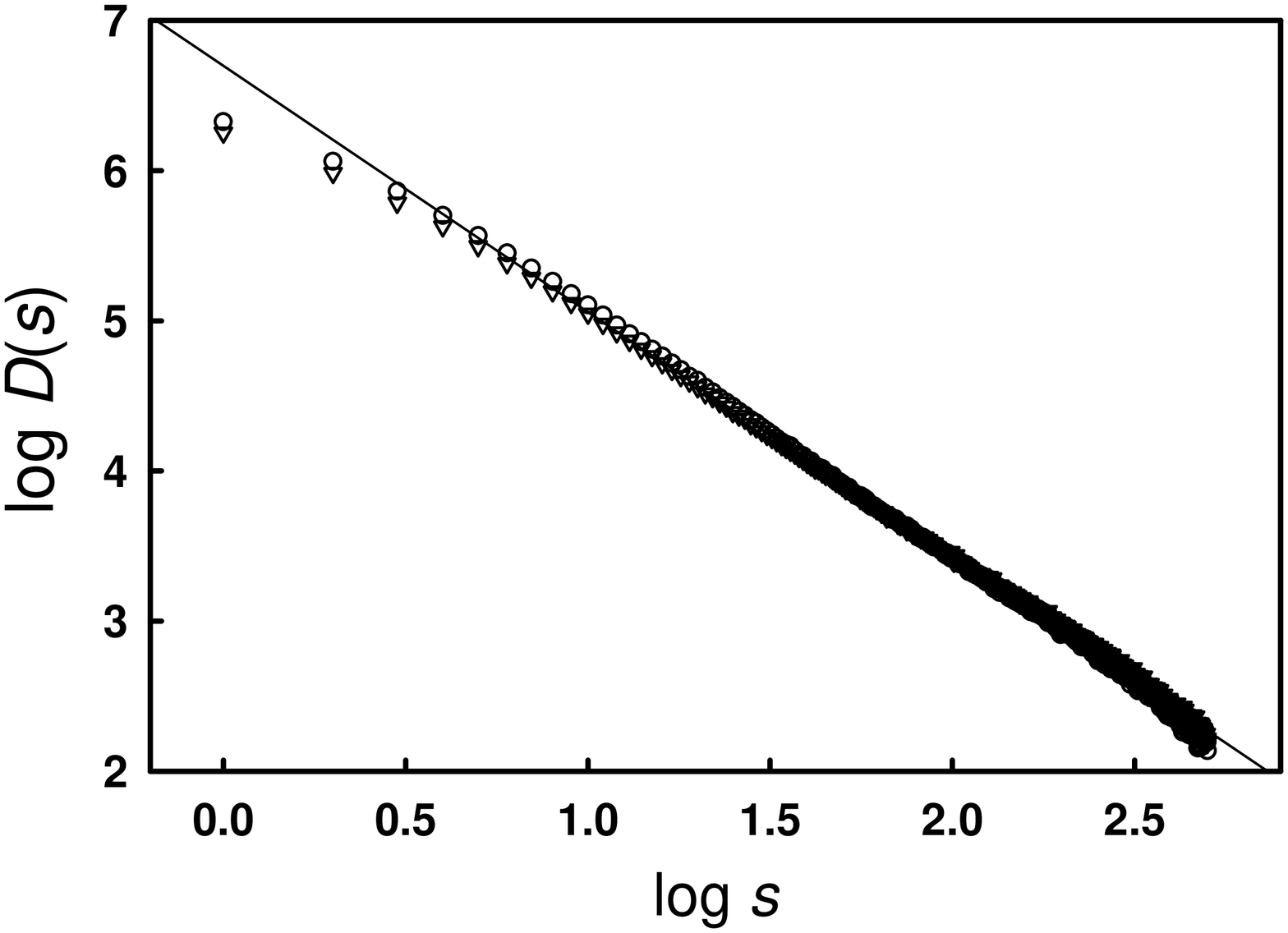}}}
\caption[0]{Avalanche size distribution $D(s)$ in the system with the fitness given by 
$f_i = x_i $ $(\circ )$ and by Eq.~(\ref{xx}) $(\triangledown)$. 
Here $D(s)$ has been obtained with $\Delta = 10$ and the straight line has 
the slope $-1.6$. 
}
\label{fig:avsd}
\end{figure}

When the fitness function is given by Eq.~(\ref{xx}) or (\ref{x-x}), the behavior
of the ecosystem appears to depend appreciably on the spatial dimension of the system.
We first present the results of the system forming a one-dimensional chain. 
For the fitness function in the form of Eq.~(\ref{xx}), the time series 
in 19 samples out of the 20 ones considered are entirely similar to those of the 
system with the BS-type fitness function presented already. 
Just one sample constitutes an exception out of the twenty, with the evolution activity 
stopping for certain periods of time; this behavior will be discussed later. 
The avalanche size distribution $D(s)$, shown in Fig.~\ref{fig:avsd}, is also similar to that 
in the system with the BS-type fitness.

\begin{figure}[b!]
\centering{\resizebox*{!}{10cm}{\includegraphics{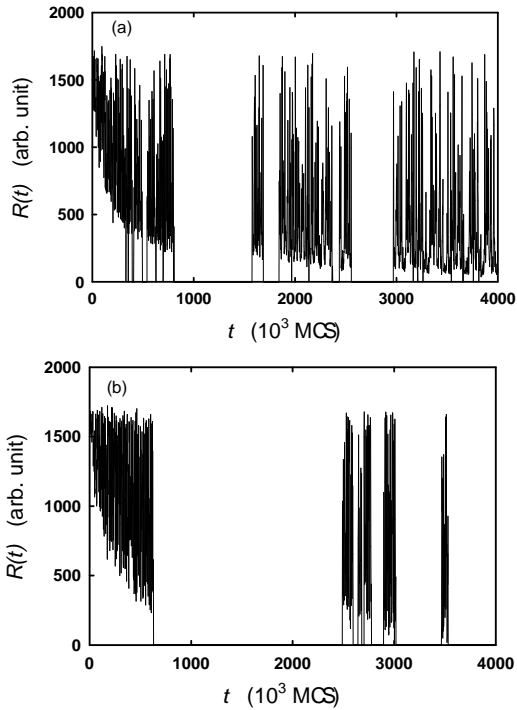}}}
\caption[0]{Coarse-grained time series $R(t)$ (a) of a one-dimensional system with
the fitness function given by Eq.~(\ref{x-x}); (b) of a two-dimensional system with the fitness
function given by Eq.~(\ref{xx}).
}
\label{fig:pause}
\end{figure}

In case that the fitness function is given by Eq.~(\ref{x-x}), only about $10\,\%$ of
the samples in the set studied evolve smoothly into the stationary state without pause. 
A typical time series, shown in Fig.~\ref{fig:pause}(a), consists of active periods
and dormant periods (``pauses''), during which mutations do not take place at all. 
This arises from the fact that the system is fallen in a low-entropy region and it takes a long time 
for the system to escape from it. 
Note that this does not happen in the case of the BS-type fitness function, for which 
there is no coupling between species.  
In the presence of coupling, the system in the state of the minimum of $R(t)$ sometimes 
goes through a kind of frustration in searching the configuration $x$ at which the escape 
is possible. 
Such behavior can be observed far less frequently in the system with the fitness given by Eq.~(\ref{xx}) 
instead of Eq.~(\ref{x-x}); the former fitness function tends to make the system 
more synergic for evolution.

We now go over to the case of a two-dimensional square lattice under periodic boundary conditions. 
In the two-dimensional system, even with the fitness in Eq.~(\ref{xx}), 
only about $20\,\%$ of the samples in the set are observed to grow smoothly
into the stationary state without pause.  Figure~\ref{fig:pause}(b) shows the typical time
series, where the pause duration is in general longer than that in the one-dimensional system with the 
same fitness and sometimes mutations may terminate forever. 
In the case of the fitness in Eq.~(\ref{x-x}), no sample has been found to evolve smoothly:
The average duration of the pause tends to be even longer and the system finally stops evolving. 
It is thus concluded that a two-dimensional system, which is more realistic in view of the real 
ecosystem, tends to have dormant periods more often than a one-dimensional system, 
presumably due to the presence of frustration. 

Finally, we briefly mention the effects of shortcuts on the evolution activity. 
Although coupling is expected to exist mostly between neighboring species in the ecosystem, 
it is also conceivable that some species interact via long-range coupling. 
Such long-range coupling may be described by the addition of shortcuts, 
making the ecosystem possess the small-world network structure~\cite{sw}. 
We have built a small-world network by adding shortcuts to the one-dimensional system 
and investigated the evolution of the system. 
It is found that the evolution in most samples tends to become a bit more active as shortcuts
are added, with the total number of avalanches increased up to $5\,\% $. 
In the systems displaying pause periods, on the other hand, the addition of shortcuts 
may enhance or weaken the activity, depending on the initial configuration and the positions of shortcuts. 

\section{Summary}

We have studied in detail the information transfer model for mass extinction, 
with three kinds of fitness function.  The behavior of the system evolving into a stationary state 
has been found to depend on the spatial dimension of the system as well as the fitness function. 
The system with the BS-type fitness in general evolves into a stationary state, irrespective 
of the initial configuration. 
It is interesting and remarkable that the mutation rate in the stationary state exhibits
a $1/f$-like power spectrum with the exponent about $1.5$, quite similar to that of fossil data. 

When interactions between neighboring species are introduced, the evolution of the system
may be either largely similar to that of the system in the absence of interactions (i.e., with the BS-type
fitness) or characterized by active periods separated by dormant ones, 
depending on the specific form of the (interacting) fitness function. 
Such pause behavior during evolution has been attributed to the presence of frustration in 
escaping from the low-entropy region.  
In the case that species are located on a two-dimensional lattice, the pause periods appear
more often regardless of the fitness function, reflecting the enhancement of frustration. 

Further, when shortcuts representing long-range interactions are added in the system, 
generating small-world network structure, the evolution behavior in general depends on
the initial configuration: The overall activity of the system may become either stronger or weaker 
in the presence of shortcuts.  In particular, in the system evolving without pause, 
the addition of shortcuts tends to induce avalanches slightly more. 

\acknowledgments
This work was supported in part by the 2003 Research Fund of University of Ulsan. 

\end{document}